\documentclass[twocolumn,pra,amssymb]{revtex4}
\usepackage{amsmath,latexsym,revsymb,graphicx,epsfig,amssymb,verbatim,mathrsfs}
\def\m{\mathcal{M}}
\def\n{\mathcal{N}}
\def\u{\mathcal{U}}
\newcommand{\h}{{\mathscr{H}}}
\newcommand{\RR}{{\mathbb{R}}}
\newcommand{\hb}{{\mathscr{B}}}
\newcommand{\qed}{{$\Box$}}
\newlength{\blank}
\newenvironment{beweis}[1][{\hspace{-\blank}}]{{\noindent\emph{Proof~{#1}.\ }}}{\hfill\qed\vskip 0.5\baselineskip}

\begin{document}

\title{A non-distillability criterion for secret correlations}

\author{Llu\'{\i}s Masanes$^1$, Andreas Winter$^{2,3}$}

\affiliation{$^1$ICFO-Institut de Ciencies Fotoniques, 08860 Castelldefels (Barcelona), Spain \\
$^2$Department of Mathematics, University of Bristol, Bristol BS8 1TW, U.K.
\\
$^3$Centre for Quantum Technologies, National University of Singapore, 2 Science Drive 3, Singapore 117542}

%\date

\begin{abstract}
Within entanglement theory there are criteria which certify that some quantum states cannot be distilled into pure entanglement. An example is the positive partial transposition criterion. Here we present, for the first time, the analogous thing for secret correlations. We introduce a computable criterion which certifies that a probability distribution between two honest parties and an eavesdropper cannot be (asymptotically) distilled into a secret key. The existence of non-distillable correlations with positive secrecy cost, also known as bound information, is an open question. This criterion may be the key for finding bound information. However, if it turns out that this criterion does not detect bound information, then, a very interesting consequence follows: any distribution with positive secrecy cost can increase the secrecy content of another distribution. In other words, all correlations with positive secrecy cost constitute a useful resource.
\end{abstract}

\maketitle

\section{Introduction}

Information theoretic cryptology started with
Shannon~\cite{shannon:secret}, and it established that secret communication relied entirely on secret key; but not until Wyner's famous wiretap paper~\cite{wyner:wiretap} was it recognized that
noise in the eavesdropper's channel can be used to establish secrecy in a communication. The secrecy capacity of what is now called the general wiretap channel was determined in~\cite{ck}. After that, again, it took some while before the distillation of key from given correlation $P_{ABE}$ between two cooperating players, Alice (A) and Bob (B), and an eavesdropper Eve (E) was considered~\cite{maurer,AC1}, in a model where the three parties share a large number of copies of the given distribution $P_{ABE}$, and Alice and Bob can freely exchange messages over an authenticated but public channel (i.e., monitored by Eve).

\medskip

Indeed, Maurer~\cite{maurer} showed that this scenario is much richer than the one of the wiretap channel. His paper posed the problem of determining the optimal secret key rate of any given distribution $P_{ABE}$, in the discrete memoryless setting of availability of asymptotically many independent samples of the distribution, and in particular the problem of deciding whether a distribution can be distilled into a secret key or not.

\medskip

There has by now been a long history of fruitful exchange of ideas between cryptography and entanglement theory (see e.g.~\cite{BDSW}), mostly relating protocols for secret key and entanglement distillation. In the quantum case, the reverse process was considerred in~\cite{BDSW}: create a quantum state from pure entanglement with maximum efficiency. Subsequently it was shown that there exist states that require a positive rate of entanglement to be created, but yield no pure entanglement at all under any distillation procedure. These states are called \emph{bound entanglement}~\cite{bound-e,VC:bound-e}. The key to show the existence of bound entanglement is the positive partial transposition criterion \cite{Peres}, which certifies that a given state is not distillable.

\medskip

This motivated Gisin and Wolf~\cite{bound:info} to speculate on the existence of \emph{bound information}, i.e.~distributions that yield no secret key under distillation but nevertheless somehow contain secrecy. They presented some candidates for bound information derived from bound entangled quantum states. Subsequently, the notion of secret key cost of a given distribution $P_{ABE}$ was formalised (under the name~\emph{information of formation})~\cite{renner:wolf:cost}. Roughly speaking, this is the minimum amount of secret bits that are necessary in order to generate $P_{ABE}$ from public communication. Latter, a single-letter formula for this quantity was
found~\cite{winter:cost}.

\medskip

Renner and Wolf~\cite{renner:wolf:cost} have shown that there can be arbitrarily large gaps between the secret key cost and the key distillation rate, thus providing evidence for the existence of bound information (see also~\cite{winter:cost}). In~\cite{3-party-boundinfo,acin:masanes} it was shown that \emph{multipartite} (i.e., more than two honest players) bound information indeed exists. But nothing is known about the existence of bound information in the bipartite case. The reason is that no criterion for non-distillability of secret correlations is known. In this paper we present the first one, which is based on the idea presented in \cite{masanes,masanes:multi}.

\section{Notation}

Key distillation and key cost are most conveniently expressed via the probability distribution from which
Alice, Bob and Eve observe samples. A generic multipartite probability distribution among the parties $AB\!\ldots$ is denoted by a non-negative vector $P_{AB\ldots}$ belonging to the $\RR$-linear space $\h_{A}\otimes\h_{B}\otimes\cdots$ which comes with a distinguished (tensor product) basis. This distinguished, ``computational'', basis of the
local space $\h_{A}$ has one element for each outcome from the alphabet of $A$. For instance, the computational basis
of a bit $\hb$ consists of the two vectors $(1,0)$ and $(0,1)$. Note that for generic alphabets we use $\h$ to denote the
vector space, but for bits (two dimensions) we reserve $\hb$.
Furthermore, to identify which party has access to the sample from a factor in the tensor product, we attach generic indices $A$, $B$ and $E$; if the space of one party consists of several alphabets, we denote them $A$, $A'$, $A''$, etc. The coefficients of $P_{AB\ldots}$ in the computational (product) basis are denoted by $P_{AB\ldots} \!(a,b,\ldots)$, and each corresponds to the probability of the event with outcomes $(a,b\ldots)$. Hence all the coefficients of $P_{AB\ldots}$ must be non-negative. Unless explicitly mentioned, we allow probability distributions $P_{AB\ldots}$ to be not normalized. 

\medskip

A general (stochastic) operation $\m$, which may be filtering (i.e., not preserving probability), is represented by a linear map with non-negative coefficients $\m:\h_1 \rightarrow \h_2$. Because we do not care about normalization, there is no additional constraint on the coefficients of $\m$, apart from non-negativity. In the case of local operations, we specify which party performs each operation by attaching an appropriate index, e.g.~$\m_{A}\n_{B}$. We omit the tensor product sign, and the identity matrix for the remaining
parties. 

\section{Preliminary results}

The secret bit fraction was introduced in \cite{lambda}, as the secrecy analog of the quantum singlet fraction, introduced in \cite{singlet fraction}.

\medskip

{\bf Definition 1 (secret bit fraction).} Suppose $P_{ABE}$ is a tripartite normalized probability distribution, where the outcomes corresponding to parties $A$ and $B$ take values on $\{0,1\}$. The secret bit fraction of $P_{ABE}$, denoted $\lambda [P_{ABE}]$, is the maximum value of $\mu$ for which a decomposition 
\begin{equation}
    P_{ABE} = \mu\, S_{AB} P_{E}' + (1-\mu)\, P_{ABE}''
\end{equation}
exists, where $P_{E}'$ and $P_{ABE}''$ are arbitrary normalized distributions, and $S$ is a secret bit shared by two parties: $S(a,b) = \frac{1}{2} \delta_{a,b}$.

\medskip

{\bf Lemma 2.} The secret bit fraction of $P_{ABE}$ can be written as
\begin{equation}\label{lambda}
    \lambda\!\left[ P_{ABE} \right]=2
    \frac{\sum_e \min\{ P_{ABE}(0,0,e),P_{ABE}(1,1,e)\}}
    {\sum_{a,b,e} P_{ABE}(a,b,e) }\ .
\end{equation}

\medskip

This is proven in \cite{lambda}. We have included the normalization factor in the denominator of (\ref{lambda}) in order not to worry about the normalization of $P_{ABE}$; in this way, the quantity $\lambda[P_{ABE}]$ makes sense irrespective of normalization of $P_{ABE}$. Note that $\lambda[P_{ABE}]=1$ means that $P_{ABE}= S_{AB}P_{E}'$, so that $P_{ABE}$ represents a \emph{secret bit} between Alice and Bob.

\medskip

{\bf Definition 3.} The maximal extractable secret bit fraction of a given distribution $P_{ABE} \in \h_A\otimes \h_B\otimes \h_E$ is
\begin{equation}\label{Lambda}
    \Lambda[P_{ABE}]=\sup_{\m_A,\n_B} \lambda[\m_A \n_B P_{ABE}]\ ,
\end{equation}
where the optimization is made over maps \mbox{$\m_A: \h_{A} \rightarrow \hb$} and \mbox{$\n_B: \h_B \rightarrow \hb$}.

\medskip

Note that the function $\lambda[ P_{ABE} ]$ is only defined for distributions $P_{ABE}$ where the alphabets of $A, B$ are $\{0,1\}$, hence, in the definition of $\Lambda$, it is important that the range of the maps $\m_A, \m_B$ is $\hb$. On the other hand, the function $\Lambda$ is defined on probability distributions $P_{ABE}$ for random variables taking values on arbitrary alphabets. The maximal extractable secret bit fraction expresses the quality of the secret bit that can be extracted from a single copy of a given distribution. If $\Lambda[P_{ABE}] =1$ then a perfect secret bit can be extracted from a single copy of $P_{ABE}$. If $P_{ABE}$ is the product of two uniformly random bits (one for each of the honest parties) and any uncorrelated information for Eve, then $\Lambda[P_{ABE}] =1/2$. Because the output of the maps $\m_A, \m_B$ can always be an independent uniform random bit, irrespectively of the input, the range of $\Lambda$ is $[1/2,1]$. It is shown in \cite{lambda} that the quantity $\Lambda$ is a secrecy monotone, and hence, constitutes a measure of the amount of secrecy contained in a given $P_{ABE}$. Additionally, there is a relation between this single-copy secrecy measure and asymptotic distillability. It is shown in \cite{lambda} that if $\Lambda[P_{ABE}] >1/2$ then $P_{ABE}$ is distillable. In what follows we rephrase the definition of distillability in terms of $\Lambda$.

\medskip

{\bf Definition 4 (Distillability).} We say that the distribution $P_{ABE}\in \h_A\otimes\h_B\otimes\h_E$ is \mbox{(secret-key-)}distillable if for each $\lambda_0\in [1/2,1)$ there exists an integer $n$ such that $\Lambda [ P_{ABE}^{\otimes n}] > \lambda_0$.

\medskip

That is, from a sufficiently large number of copies of $P_{ABE}$, Alice and Bob can, by local operations and public communication (which, without loss of generality, can be assumed to be a filtering of the form written in (\ref{Lambda})), obtain arbitrarily good approximations to a secret bit. If there exists $n$ such that $\Lambda[P_{ABE}^{\otimes n}] >1/2$, one can apply advantage distillation~\cite{maurer} to the result and obtain a secret key (see \cite{lambda}). Note furthermore that in this case even positive rates of secret key can be obtained, as $n\rightarrow\infty$. (The reader familiar with entanglement theory will realize the similarity of these concepts to singlet fraction and singlet distillability.) The difficulty in dealing with distillability is that its definition involves an arbitrarily large number of copies of $P_{ABE}$. The following tools deal with this problem.

\medskip

{\bf Lemma 5.} Let $\h_1$, $\h_2$ and $\h_3$ be given vector spaces. Any ``global'' linear map $\m: \h_1 \otimes \h_2 \rightarrow \h_3$ with \mbox{non-negative} coefficients can be decomposed into a local linear map with non-negative coefficients, $\m': \h_1 \rightarrow \h_3 \otimes \h_2$ (which depends on $\m$), and a simple global linear map with non-negative coefficients, $\u: (\h_3\otimes \h_2)\otimes \h_2 \rightarrow \h_3$ (which is independent of $\m$, that is, universal, and given by $\u^{y_3}_{x_3 x_2 y_2} = \delta^{y_3}_{x_3}\, \delta_{x_2 y_2}$), such that $\m=\u\m'$.

\medskip

\begin{beweis}
If we adopt the convention that lower indices correspond to the input and upper indices to the output, we can write $\m'$ in terms of $\m$ as $\m'^{x_3 x_2}_{x_1} = \m^{x_3}_{x_1 x_2}$. The equality 
\begin{equation}\label{}
  \m^{y_3}_{x_1 y_2}= \sum_{x_2 x_3 y_2'} 
  \u^{y_3}_{x_3 x_2 y_2'} \big( \m'^{x_3 x_2}_{x_1} 
  \delta^{y_2'}_{y_2} \big)\ ,
\end{equation}
holds by definition.
\end{beweis}

\medskip

{\bf Lemma 6.} If the distribution $P_{ABE}\in$
\mbox{$\h_A\otimes\h_B\otimes\h_E$} is distillable, then for each $\lambda_0\in [1/2,1)$ there exists a distribution $Q_{ABE'} \in (\hb_{A}\otimes\h_{A}) \otimes (\hb_{B}\otimes\h_{B}) \otimes\h_{E'}$  such that
\begin{eqnarray}\label{i1}
    \Lambda[Q_{ABE'}] &\leq& \lambda_0\ ,
    \\ \label{i2}
    \lambda\!\left[ \u_A \u_B\, Q_{ABE'}\otimes P_{ABE} \right]
    &>& \lambda_0\ ,
\end{eqnarray}
where $\u$ is defined in Lemma 5. The size of $\h_{E'}$ is arbitrary.

\medskip

\begin{beweis}
Let $n$ be the smallest integer such that there exist operations $\m_A:\h_A^{\otimes n} \rightarrow \hb_A$ and $\n_B:\h_B^{\otimes n} \rightarrow \hb_B$ such that $\lambda [ \m_A \n_B \, P_{ABE}^{\otimes n} ] > \lambda_0$ (following Definition 4). According to Lemma 5 there are maps $\m_A', \n_B'$ such that $\m_A = \u_A\m_A'$, $\n_B = \u_B\n_B'$, and the distribution $Q_{ABE'} = \m_A' \n_B' \, P_{ABE}^{\otimes(n-1)}$ has alphabet $(\hb_{A}\otimes\h_{A}) \otimes (\hb_{B}\otimes\h_{B}) \otimes\h_{E'}$, as we want to show. Because $\Lambda$ is defined through an optimization (Definition 3), we have
\begin{equation}\label{cad}
  \Lambda[P_{ABE}^{\otimes(n-1)}] \geq \Lambda[M_A' \n_B' \, P_{ABE}^{\otimes(n-1)}] = \Lambda[Q_{ABE'}]\ .
\end{equation}
The definition of $n$ implies that $\Lambda[P_{ABE}^{\otimes(n-1)}] \leq \lambda_0$, which together with (\ref{cad}), implies (\ref{i1}). Using the properties of the maps $\u,\m',\n'$ shown in Lemma 5, one can check that
\begin{equation}\label{8}
  \u_A \u_B\, Q_{ABE'}\otimes P_{ABE} = \m_A \n_B \, P_{ABE}^{\otimes n}\ .
\end{equation}
Recall that the maps $\m_A,\n_B$ are the ones for which $\lambda [ \m_A \n_B \, P_{ABE}^{\otimes n} ] > \lambda_0$, which together with (\ref{8}), implies (\ref{i2}).
\end{beweis}

\medskip

In other words, what Lemma 6 tells is that if a distribution $P_{ABE}$ is distillable, then it can activate the secrecy of another distribution $Q_{ABE}$. Here by activation we mean enhancement of the maximal extractable secret bit fraction $\Lambda[\cdot]$. The important point is that Alice's and Bob's alphabets in $Q_{ABE}$ are bounded. Unfortunately, Lemma 6 does not tell anything about the size of Eve's alphabet in $Q_{ABE'}$, that is $\h_{E'}$, but this problem will later sort out itself.

\section{Non-distillability criterion}

In order to certify that a given distribution $G_{ABE} \in \h_A\otimes \h_B\otimes \h_{E}$ is undistillable, it suffices to obtain a contradiction between the inequalities (\ref{i1}) and (\ref{i2}). However, the characterization of the set of distributions $Q_{ABE'}\in (\hb_{A}\otimes\h_{A}) \otimes (\hb_{B}\otimes\h_{B}) \otimes\h_{E'}$, where the size of $\h_{E'}$ is arbitrary, satisfying $\lambda[\m_A \n_B Q_{ABE'}] \leq \lambda_0$ for any pair of maps $\m_A, \n_B$ is not available. Instead, we consider a larger (but simpler) set. For any given finite family of pairs of maps ${\cal F} = \{(\m_A^i, \n_B^i) : i= 1,\ldots M \}$, we consider the set of distributions which satisfy $\lambda[\m_A^i \n_B^i Q_{ABE'}] \leq \lambda_0$ for $i=1,\ldots M$. In what follows we particularize to $\lambda_0 =1/2$, although different criteria could be obtained for different values of $\lambda_0$. Another big simplification is to write the inequalities (\ref{i1}) and (\ref{i2}) as ``almost"-linear in the vector $Q_{ABE'}$. If we denote by $e$ the variable of $\h_E$, and by $e'$ the variable of $\h_{E'}$, we can write (\ref{i1}) and (\ref{i2}) as
\begin{align}
    &2\sum_{e',e} \min_{a\in \{0,1\}}\!
    \Big\{[\u_A\u_B\, Q_{ABE'}\otimes G_{ABE}](a,a,e',e)\Big\}    
    \nonumber \\
    &\quad -\frac{1}{2}\sum_{a,b,e',e} [\u_A \u_B\, Q_{ABE'} \otimes G_{ABE}](a,b,e',e) >0
    \label{li1} \\
    &2\sum_{e'} \min_{a\in \{0,1\}}\!
    \Big\{[\m^i_A \n^i_B\, Q_{ABE'}](a,a,e')\Big\}
    \nonumber \\
    &\quad -\frac{1}{2}\sum_{a,b,e'}\ [\m^i_A \m^i_B\, Q_{ABE'}](a,b,e')   \leq 0\ , \label{li2}
\end{align}
for $i=1,\ldots M$. This is obtained by using the explicit form of $\lambda[\cdot]$ given in (\ref{lambda}), and setting $\lambda_0 =1/2$.

\medskip

Denote by $d$ the dimension of $\h_E$. In (\ref{li1}) and (\ref{li2}) the summation over $e$ runs over $d$ values, while the summation over $e'$ is unbounded (like the dimension of $\h_{E'}$). In what follows we transform the summation over $e'$ into one over $2^{d+M}$ values. For each $e=1,\ldots d$, define the function
\begin{equation}\nonumber
  r_e(e') = \left\{
  \begin{array}{lll}
    0 &\mathrm{if} &\sum_{a} (-1)^{a} [\u_A \u_B Q_{ABE'}\otimes G_{ABE}]
    (a, a, e', e) <0
    \\
    1 &\mathrm{if} &\sum_{a} (-1)^{a} [\u_A \u_B Q_{ABE'}\otimes G_{ABE}]
    (a, a, e', e) >0
  \end{array}
  \right.
\end{equation}
for all $e'$. Analogously, for each $i=1,\ldots M$ define the function
\begin{equation}\nonumber
  s_i(e') = \left\{
  \begin{array}{lll}
    0 &\mathrm{if} &\sum_{a} (-1)^{a} [\m^i_A \n^i_B Q_{ABE'}]
    (a, a, e') <0
    \\
    1 &\mathrm{if} &\sum_{a} (-1)^{a} [\m^i_A \n^i_B Q_{ABE'}]
    (a, a, e') >0
  \end{array}
  \right.
\end{equation}
for all $e'$. Using these definitions we can write, for any value of $e,i,e'$,
\begin{align}
    &\min_{a\in \{0,1\}}
    \Big\{[\u_A \u_B\, Q_{ABE'}\otimes G_{ABE}](a,a,e',e)\Big\}
    \nonumber \\ \label{k1}
    &\quad =[\u_A \u_B\, Q_{ABE'}\otimes G_{ABE}](r_e(e'), r_e(e'),e' ,e)\ ,
\\
    &\min_{a\in \{0,1\}}
    \Big\{[\m^i_A \m^i_B\, Q_{ABE'}](a,a,e')\Big\}
    \nonumber \\ \label{k2}
    &\quad =[\m^i_A \m^i_B\, Q_{ABE'}](s_i(e'), s_i(e') ,e')\ ,
\end{align}
which allows to get rid of the $\min$-functions in (\ref{li1}) and (\ref{li2}). Let us define the new variable ${\bf k}$ in the following way
\begin{equation}
  {\bf k}(e') = (r_0 (e'), r_1(e'), \ldots r_d(e'),s_1(e'),\ldots s_M(e')),
\end{equation}
which has the natural distribution and correlations with $A,B$,
\begin{equation}\label{K}
  Q_{ABK} (a,b,{\bf k}_0) = \sum_{e': {\bf k}(e') ={\bf k}_0} Q_{ABE'} (a,b,e')\ .
\end{equation}
This allows to write the identities
\begin{align}
    &\sum_{e',e} \min_{a\in \{0,1\}}
    \Big\{[\u_A \u_B\, Q_{ABE'}\otimes G_{ABE}](a,a,e',e)\Big\}
    \nonumber \\ \label{k1}
    &\quad = \sum_{{\bf k},e}\ [\u_A \u_B\, Q_{ABK}\otimes G_{ABE}](k_e, k_e, {\bf k},e)\ ,
\\
    &\sum_{e'} \min_{a\in \{0,1\}}
    \Big\{[\m^i_A \m^i_B\, Q_{ABE'}](a,a,e')\Big\}
    \nonumber \\ \label{k2}
    &\quad = \sum_{\bf k}\ [\m^i_A \m^i_B\, Q_{ABK}](k_{d+i}, k_{d+i},{\bf k})\ ,
\end{align}
where we have used the fact that when $\{ x_0\leq x_1$ and $y_0\leq y_1 \}$ or \mbox{$\{ x_0\geq x_1$ and $y_0\geq y_1 \}$} the equality
\begin{equation}\label{grouping}
    \min\{x_0,x_1\}+\min\{y_0,y_1\}
    =\min\{x_0+y_0,x_1+y_1\}
\end{equation}
holds. After grouping the different values of $e'$ as in (\ref{K}), we only need to consider distributions $Q_{ABK}$ where the variable ${\bf k}$ runs over $2^{d+M}$ different values. However, the new (bounded in size) distribution $Q_{ABK}$ must satisfy the constraints
\begin{align}
    &\sum_{a} (-1)^{a} [\u_A \u_B Q_{ABK}\otimes G_{ABE}]
    (k_e \oplus a, k_e \oplus a, {\bf k}, e) <0\ ,
    \nonumber \\ \nonumber
    &\sum_{a} (-1)^{a} [\m^i_A \m^i_B Q_{ABK}]
    (k_{d+i} \oplus a, k_{d+i} \oplus a, {\bf k}) <0\ ,
\end{align}
for all $e,i,{\bf k}$.

\medskip

Now everything is finite. $Q_{ABK}$ is a vector from the space $(\dim\h_A \times \dim\h_B \times 2^{d+M+2})$ with non-negative components, that is $Q_{ABK}(a,b,{\bf k}) \geq 0$ for all $a,b, {\bf k}$. Hence, the set of allowed distributions $Q_{ABK}$ is characterized by a finite set of linear inequalities. Then, maximizing the left-hand side of (\ref{li1}) is a linear programming problem.
\begin{align}\label{lp} 
    &\mbox{LINEAR PROGRAMMING:}
    \\ \nonumber
    &\mbox{[If the maximum is zero then $G_{ABE}$ is undistillable.]}
    \\ \nonumber
    &\max_{Q_{ABK}} \sum_{{\bf k},e} \Big( 4
    [\u_A \u_B\, Q_{ABK}\otimes G_{ABE}] (k_e,k_e,{\bf k},e)-      
    \\ \nonumber
    &\quad -\sum_{a,b} [\u_A \u_B\, Q_{ABK} \otimes G_{ABE}] (a,b,{\bf k},e) \Big)
    \\ \nonumber &\mbox{with constrains}
    \\ \nonumber
    &4\sum_{\bf k}
    [\m^i_A \m^i_B\, Q_{ABK}](k_{d+i},k_{d+i},{\bf k})-
    \\ \nonumber
    &\quad -\sum_{a,b,{\bf k}}\ [\m^i_A \m^i_B\, Q_{ABK}] (a,b,{\bf k})   \leq 0\ ,
    \\ \nonumber
    &\sum_{a} (-1)^{a} [\u_A \u_B\, Q_{ABK}\otimes G_{ABE}]
    (k_e \oplus a, k_e \oplus a, {\bf k}, e) <0\ ,
    \\ \nonumber
    &\sum_{a} (-1)^{a} [\m^i_A \m^i_B\, Q_{ABK}]
    (k_{d+i} \oplus a, k_{d+i} \oplus a, {\bf k}) <0\ ,
    \\ \nonumber
    &\sum_{a,b,{\bf k}} Q_{ABK}(a,b,{\bf k}) =1
    \\ \nonumber
    &\ Q_{ABK}(a,b,{\bf k}) \geq 0\ ,
\end{align}
for all $i=1,\ldots M$, all ${\bf k} \in \{0,1\}^{d+M}$, and all $a,b \in \{0,1\}$ in the last inequality.

\medskip

If the given distribution $G_{ABE}$ has rational coefficients, the above linear programming can be solved by exact methods like the simplex algorithm~\cite{LP}. Or by quasi-exact methods like the interior point algorithm \cite{BV}, whose solution can always be certified exactly. The last method is faster, and hence can deal with larger values of $M$.

\medskip

A key feature of this method is to choose a suitable family $\mathcal{F}$ of pairs of maps. The larger the size of this family ($M$) the more constrains on the above maximization, and more chances to get the maximum equal to zero.

\section{remarks}

If the maximum of the linear programming (\ref{lp}) is zero then we know for sure that $G_{ABE}$ is undistillable. But actually, we know  something much stronger: $G_{ABE}$ cannot activate any other non-distillable distribution. In other words, the correlations in $G_{ABE}$ are completely useless.

\medskip

{\bf Lemma 7.} {\em Let the distribution $G_{ABE}$ be such that the maximum of the linear programming (\ref{lp}) is zero. If $P_{ABE}$ is a non-distillable distribution, then the tensor-product $P_{ABE}\otimes G_{ABE}$ is also non-distillable.}

\medskip

\begin{beweis}
  By assumption, for any distribution $Q_{ABE}$ such that $\Lambda[Q_{ABE}] \leq 1/2$ we have
  $\Lambda[Q_{ABE} \otimes G_{ABE}] \leq 1/2$. In particular, if we chose $Q_{ABE} = P_{ABE}^{\otimes n}$ we have $\Lambda[P_{ABE}^{\otimes n} \otimes G_{ABE}] \leq 1/2$, for any $n$. But this also implies $\Lambda[(P_{ABE}^{\otimes n} \otimes G_{ABE}) \otimes G_{ABE}] \leq 1/2$, and proceeding by induction, we obtain $\Lambda[(P_{ABE} \otimes G_{ABE})^{\otimes n}] \leq 1/2$.
\end{beweis}

An interesting possibility is that for any distribution $G_{ABE}$ with positive secrecy cost  all linear programming problems (\ref{lp}) have a larger than zero maximum. This would imply that our criterion does not detect any non-distillable distribution, and hence it is useless. But this would also imply that any distribution $G_{ABE}$ with positive secrecy cost (even though it may be non-distillable) can increase the quality of the secret bits distilled from a single copy of another distribution $Q_{ABE}$. Actually, an analog of the last statement is true in the quantum case \cite{masanes, masanes:multi}. That is, all entangled states (of any number of parties) can increase the quality of the entanglement that can be distilled from a single copy of another state.

\section{Conclusion}

In this paper, we have presented the first criterion which certifies that a given distribution $G_{ABE}$ has no distillable key. In fact, this method consists on showing that $G_{ABE}$ does not improve the key content of \emph{any} other distribution (i.e., it does not bring the maximally extractable secret bit fraction above $1/2$). 

\medskip

It is an open question whether all correlations with positive secrecy content can increase the secrecy of other correlations. This very interesting feature of secret correlations would invalidate the criterion presented here.

\medskip

Finally, and perhaps most interestingly: does our technique have a quantum analogue, which could be used to prove the existence of entangled quantum states that do not contain secret key? This would present a complement to the work by Horodecki et al.~\cite{HHHO-05}, who show the existence of bound entangled states that do nevertheless contain secret key: perhaps there exists \emph{completely bound entanglement} which neither contains distillable key nor enhances key content of other states.

\medskip

{\bf Acknowledgments.} LlM is supported by the spanish MEC (FIS2005-04627, FIS2007-60182, Consolider QOIT), and Caixa Manresa. AW is supported by the U.K. EPSRC (project "QIP IRC" and an Advenced Research Fellowship), and by a Royal Society Wolfson Merit Award.

\end{document}